\documentclass{moriond}

\usepackage{graphicx,bm,amsmath,amssymb,latexsym,psfrag,epsfig,euscript, multirow,color,verbatim,cancel}
\usepackage[utf8]{inputenc}

\def\be{\begin{equation}}
\def\ee{\end{equation}}
\def\bea{\begin{eqnarray}}
\def\eea{\end{eqnarray}}

\newcommand{\Photo}{}

\begin{document}
\vspace*{4cm}
\title{Standard Model -- Axion -- Seesaw -- $H$ portal inflation}

\author{Carlos Tamarit\footnote{E-mail:carlos.tamarit@durham.ac.uk}}

\address{Institute for Particle Physics Phenomenology, Durham University, DH1 3LE, United Kingdom.}

\maketitle\abstracts{
Extending the Standard Model with a new complex singlet scalar, right-handed neutrinos and a vector-like quark allows to simultaneously tackle several problems in particle physics and cosmology within a constrained framework
that can be falsified by future probes of the cosmic microwave background, as well as by upcoming axion experiments.
This Standard  Model - Axion - Seesaw - $H$ portal inflation theory (SMASH) provides predictive inflation 
and $H$ boson stabilization, and can explain baryogenesis, light neutrino masses, dark matter and the strong CP problem. The model contains a unique new mass scale which coincides with the axion decay constant, and also sets the scale for perturbative lepton-number violation processes. Testable predictions include a minimum value of the tensor-to-scalar ratio of $r\gtrsim 0.004$, a running of the spectral index  $\alpha\gtrsim-8\times10^{-4}$, a change $\delta N_{\rm eff}\sim 0.03$ in the  number of effective relativistic neutrinos, and an axion mass in the range $50\mu eV\leq m_A \leq 200 \mu eV$.
}

\section{Minimal extensions of the Standard Model}

The results of experiments on Earth and in space are vastly compatible with the theoretical framework obtained by, on the one hand, extending the Standard Model of particle physics (SM)  to accommodate for neutrino masses, and on the other by assuming some form of cold dark matter in the Universe, together with early and late periods of accelerated expansion driven, respectively, by inflationary dynamics and the cosmological constant. Despite its success, 
the framework does not provide unique answers to questions such as the microscopic origin of neutrino masses,  dark matter and inflation. Furthermore, the Standard Model alone cannot explain 
the origin of the observed baryon asymmetry (the Universe is overwhelmingly made of matter, rather than anti-matter), and is still plagued by puzzles such as the apparent absence of CP violation
in the strong interactions (strong CP problem), and the  instability that appears in the $H$ boson's potential, for the preferred values of the $H$  and top quark masses, when the field becomes large \cite{Degrassi:2012ry,Bednyakov:2015sca}. The latter problem
is particularly troublesome when considering its interplay with inflation and the ensuing reheating of the Universe, as both processes can generate large perturbations in the $H$ boson that drive it towards the unstable region, preventing it from reaching the electroweak vacuum \cite{Espinosa:2007qp}. Other well-known problems, which won't be addressed here, are related to the naturalness of the electroweak and cosmological constant scales.

Each of the  problems  of inflation, $H$ stability, baryogenesis, neutrino masses, dark matter and strong CP have known solutions that can be realized in terms of new particles and interactions. For example, inflation can be sourced by the dynamics of a slowly-rolling scalar field with a positive energy density. The $H$ particle can be stabilized by means of new bosonic interactions. Baryogenesis may proceed from CP violation in out-of-equilibrium decays of heavy particles, or it can be created in first-order phase transitions, or as a byproduct of some other scalar field dynamics. Neutrino masses can be explained through a seesaw mechanism, triggered by new heavy particles, or by some radiative effect. Dark matter can be a new fermion or boson which acquires the observed relic density by either thermal or non-thermal dynamics, and the strong CP problem can be addressed with axions, or assuming a spontaneously broken exact CP symmetry.  When considering extensions of the Standard Model which realize the previous solutions, one may implement them with independent sectors, or one may try to find more minimal setups in which the different solutions become intertwined. In this spirit, one may aim for minimal models addressing as many solutions as possible. There is a long history in the literature of combined solutions and minimal models. A model that stands out for its minimality is the $\nu{\rm MSM}$ \cite{Asaka:2005an,Asaka:2005pn}. By simply adding three right-handed neutrinos $N_i$ to the Standard Model, the $\nu{\rm MSM}$ can provide inflation through the $H$ scalar field \cite{Bezrukov:2007ep} (with a non-minimal gravitational  coupling), light neutrino masses through a seesaw mechanism \cite{Minkowski:1977sc} involving the $N_i$, baryogenesis through flavoured oscillations of $N_2,N_3$ (required to have GeV scale masses) \cite{Akhmedov:1998qx}, and dark matter from a keV scale $N_1$. Despite these remarkable properties, $H$ inflation in the $\nu{\rm MSM}$ has been argued to lack predictivity, due to problems with unitarity 
\cite{Burgess:2009ea,Barbon:2009ya}. Moreover, it requires a stabilized $H$ potential at large field values, which the model fails to provide for the current central values of the $H$ and top masses (although stability is not yet  ruled out \cite{Degrassi:2012ry,Bednyakov:2015sca}). Other minimal models addressing a host of problems in particle physics and cosmology, which also include right-handed neutrinos, are for example the new minimal Standard Model of reference \cite{Davoudiasl:2004be}, (which relies on two extra scalars for inflation and dark matter, and fails to stabilize a 125 GeV $H$ particle) and the model of reference \cite{Salvio:2015cja}, which is similar to the $\nu{\rm MSM}$ (and shares its problems with $H$ inflation) but includes an added complex scalar and vector-like quark which implement a KSVZ axion solution to the strong CP problem \cite{Peccei:1977hh,Weinberg:1977ma,Wilczek:1977pj,Kim:1979if,Shifman:1979if}, with the axion providing dark matter \cite{Preskill:1982cy,Abbott:1982af,Dine:1982ah}, and with baryogenesis proceeding through thermal leptogenesis \cite{Fukugita:1986hr}.

Within this theory landscape, the SMASH model described here, which was put forth in references \cite{Ballesteros:2016euj} and \cite{Ballesteros:2016xej} (building upon the ideas in \cite{Dias:2014osa}), is similar in spirit to the theory of reference \cite{Salvio:2015cja}, with two important model-building differences: the $H$ is released from the task of single-handedly inflating the Universe, which avoids unitarity problems, and the right-handed neutrino and axion sectors are related by a single new physics scale.  This makes SMASH highly predictive. In our work we undertook detailed analyses of how the Universe evolves according to the model, how it addresses the problems mentioned earlier, and worked out its main falsifiable predictions. In SMASH, inflation arises from the combined dynamics of the $H$ scalar and a new complex singlet $\sigma$, which also stabilizes the $H$ direction. The field $\sigma$ is charged under an anomalous Peccei-Quinn (PQ) symmetry (with the anomaly arising due to the existence of a new vector-like quark), and its vacuum expectation value (VEV) sets the scale of right-handed neutrino masses, which can give rise to baryogenesis through thermal leptogenesis, and also explain light neutrino masses through the seesaw mechanism. The phase of $\sigma$ provides an axion field that can explain the dark matter of the Universe and solve the CP problem thanks to the PQ anomaly, like in the KSVZ model.

\section{The SMASH theory}

As the $\nu{\rm MSM}$, the SMASH theory includes right-handed neutrinos $N_i$ (at least two; we will assume three in order to allow all light neutrinos to be massive).  There is also a complex scalar $\sigma$, a singlet under the SM gauge group, and a vector-like quark made of two Weyl spinors $Q$ and $\tilde Q$ in the fundamental and anti-fundamental representation of the colour group, respectively. $Q,\tilde Q$ are assumed to have hypercharges $\mp1/3$ (or $\pm2/3$) in order to allow for a mixing with down (up) quarks and prevent the existence of stable exotic charged particles, which are 
severely constrained by experiments \cite{Chuzhoy:2008zy}. We assume a U(1) PQ symmetry under which $\sigma$ has unit charge, and which remains anomalous under SU(3) thanks to the $Q,\tilde Q$. The non-zero charges are given in table \ref{tab:PQ}, using Weyl-spinor notation. The most general scalar potential compatible with the symmetries of the model can be written as
\begin{equation}
\label{eq:V}
V(H,\sigma )= \lambda_H \left( H^\dagger H - \frac{v^2}{2}\right)^2
+\lambda_\sigma \left( |\sigma |^2 - \frac{v_{\sigma}^2}{2}\right)^2+
2\lambda_{H\sigma} \left( H^\dagger H - \frac{v^2}{2}\right) \left( |\sigma |^2 - \frac{v_{\sigma}^2}{2}\right)\,,
\end{equation}
while the  Yukawa interactions of the Weyl fermions are (with $i,j$ labelling the three generations, and choosing $\mp1/3$ hypercharges for $Q,\tilde Q$)
\begin{equation}
\begin{aligned}
  {\cal L}\supset&-\Bigg[{Y_u}_{ij}q_i\epsilon H u_j+{Y_d}_{ij}q_i H^\dagger d_j+G_{ij} L_i H^\dagger E_j + F_{ij}L_i\epsilon H N_j+\frac{1}{2}Y_{ij}\sigma N_i  N_j \\
 &
+y\, \tilde Q \sigma Q+\,{y_{Q_d}}_{i}\sigma Q d_i+h.c.\Bigg]\,.
\label{lyukseesaw}
\end{aligned}
\end{equation}
Note how the symmetry forbids a tree-level mass term for the $N_i$ and for $Q,\tilde Q$, whose mass is set by $\langle\sigma\rangle$. 
\begin{table}[t]
\centering
\begin{equation}{\nonumber
\begin{array}{|c|c|c|c|c|c|c|c|c|}
\hline
  q & u & d & L & N & E    & Q &\tilde Q & \sigma  \\
\hline
 1/2 & -1/2 & -1/2 & 1/2 & -1/2 & -1/2   & -1/2 & -1/2 &1 
\\[.5ex]
\hline
\end{array}
}
\end{equation}
\caption{\small Nonzero charges of the Weyl spinor and scalars under the $U(1)$ PQ symmetry in SMASH.
\label{tab:PQ}}
\end{table}

\section{SMASHy history of the Universe}

\begin{figure}[h!]
\begin{center}
\includegraphics[width=0.9\textwidth]{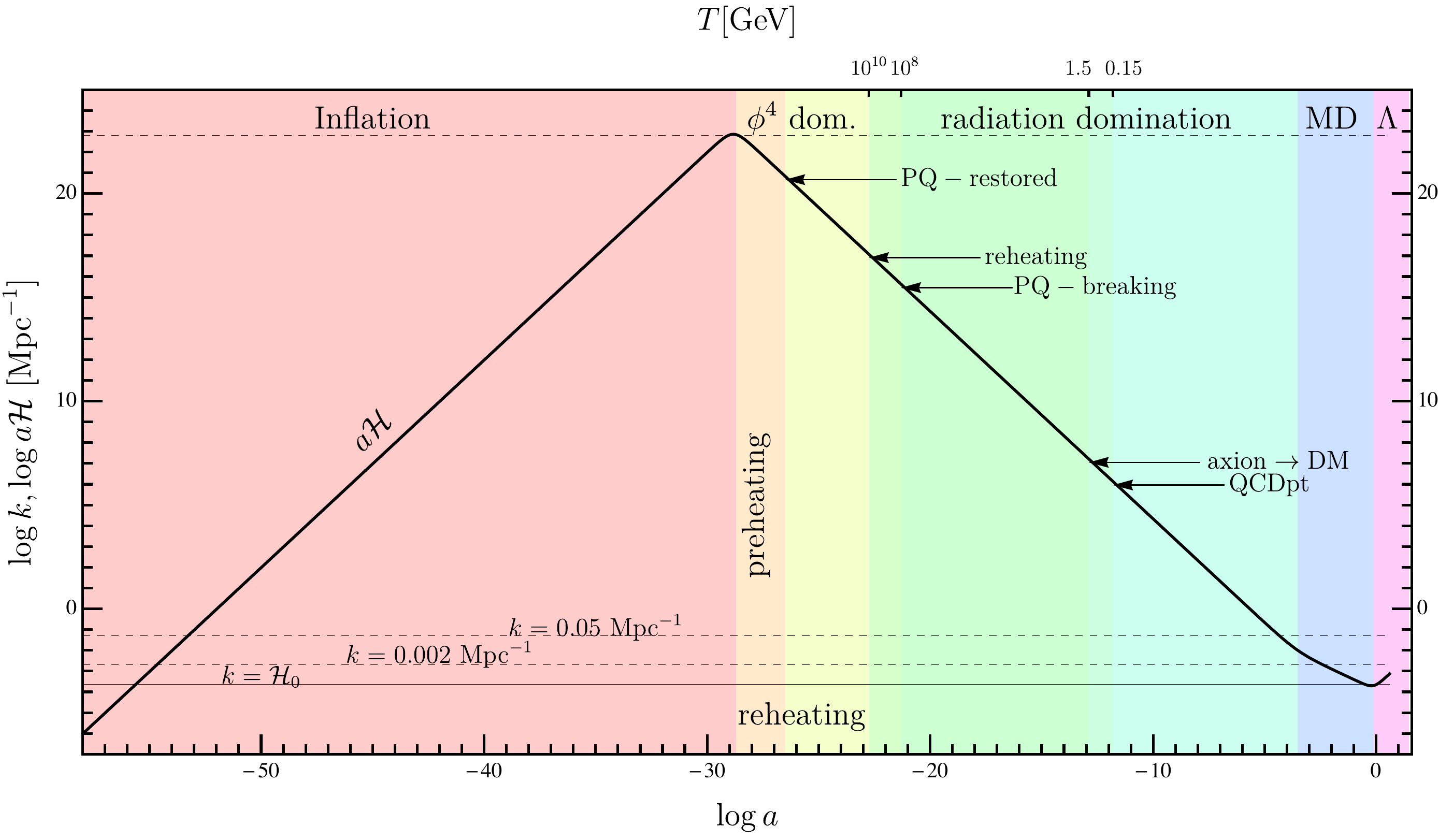}
\caption{(Taken from$^{23}$). SMASHy history of the Universe, illustrated by following the evolution of the inverse  of the horizon length scale, $a(t)\cal{ H}$,  with $a(t)$ denoting the scale factor of the Universe, and  $\cal{ H}$ the Hubble expansion rate.}  
\label{fig:SMASHy}       
\end{center}
\end{figure}

The SMASH model provides a well-defined picture of particle physics from the electroweak to the Planck scale, and gives rise to a predictive cosmological history of the Universe from 
inflation until the present. The evolution of the Universe according to SMASH is as follows (see figure  \ref{fig:SMASHy}, taken from \cite{Ballesteros:2016xej}, for a summary): a period of inflation driven by  $\sigma$ with a small admixture of $H$ is followed by a preheating phase in which the scalar degrees of freedom involved in inflation oscillate in a quartic potential, behaving as a radiation fluid. This opens up a period of radiation domination shortly after inflation, which continues as the Universe
reheats into SM degrees of freedom. The PQ symmetry is restored by non-thermal and thermal effects,  eventually breaking as the Universe cools down. The RH neutrinos acquire then a mass and subsequently decay producing the baryon asymmetry of the Universe. Once the electroweak symmetry becomes broken at lower temperatures, the light neutrinos acquire small masses through the seesaw mechanism. As the cooling proceeds
and the QCD interactions become strong, they generate a mass for the phase of $\sigma$ (the axion field), which starts oscillating and behaves as dark matter, while the effective CP violating angle in the strong interactions is relaxed to zero. After that, the evolution of the Universe is standard, with a late period of matter domination followed by the current accelerated expansion. 

\subsection{Inflation and stability\label{subsec:inflst}}

\begin{figure}[t]
\begin{center}
\includegraphics[width=0.8\textwidth]{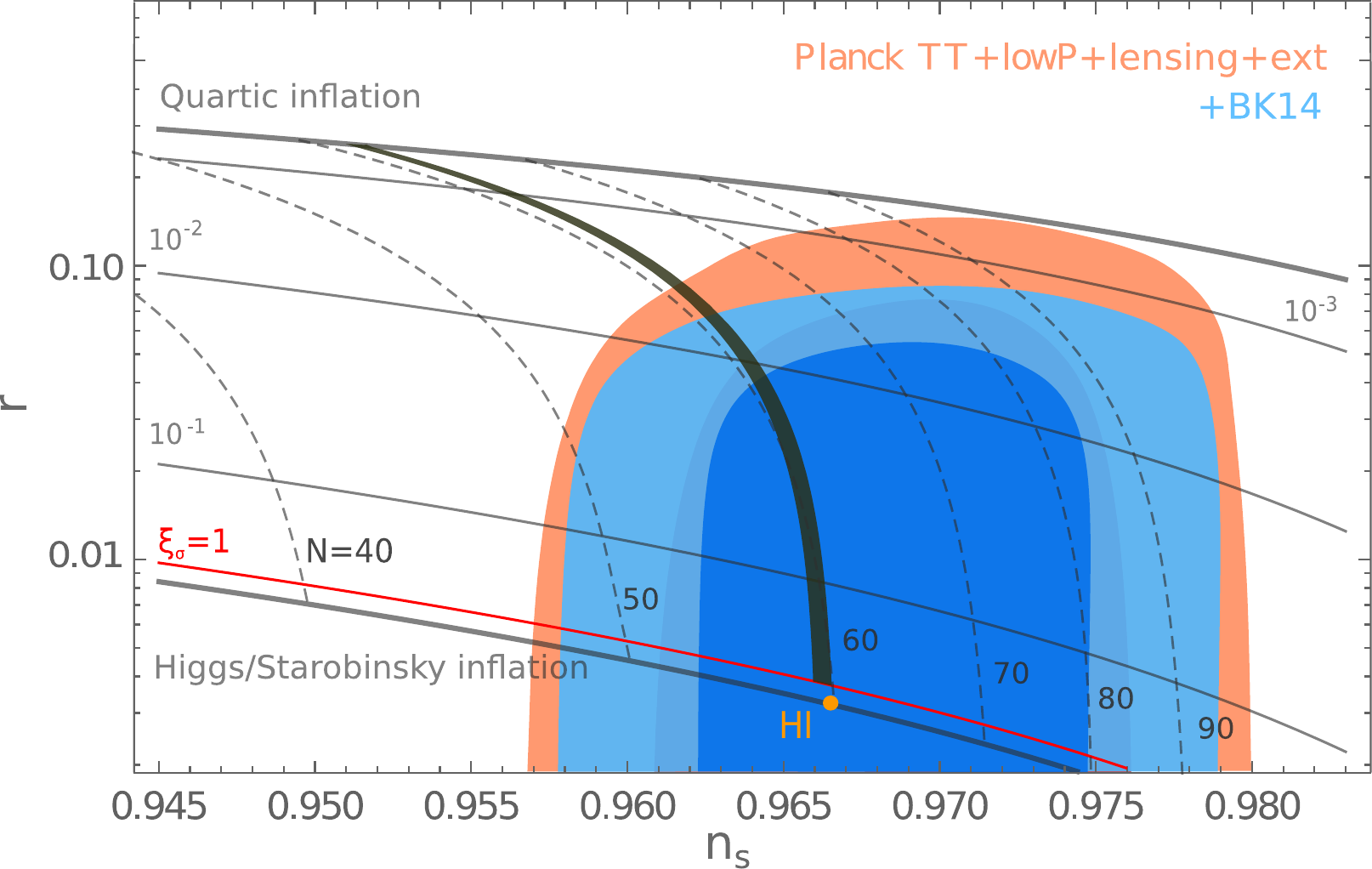}
\caption{\label{fig:CMB}  \small 
{(From$^{23}$). Inflationary predictions for $r$ and $n_{s}$ (at a scale of  0.002 Mpc$^{-1}$) in SMASH, where the thick black line takes into account the post-inflationary history.}
}
\end{center}
     \end{figure}

The scalar potential of equation \eqref{eq:V} supports slow-rolling solutions with a positive energy density, giving rise to inflation. The potential energy valleys will act as attractors for inflation, allowing to consider a one-dimensional effective dynamics. Although the latest Planck data discard inflation from a minimally coupled scalar with monomial interactions, one can obtain viable models by considering non-minimal gravitational couplings to the Ricci curvature $R$ of the form
\begin{equation}
  \label{eq:Lnm}
  S\supset - \int d^4x\sqrt{- g}\,\left[
     \frac{M^2}{2}  + \xi_H\, H^\dagger H+\xi_\sigma\, \sigma^* \sigma  
  \right] R
  \,.
\end{equation}
By performing a Weyl rescaling of the metric, the theory becomes equivalent to that of minimally coupled scalars with a modified potential. This potential has again inflationary valleys, along which the effective dynamics can be described in terms of a canonically normalized field $\chi$ with a potential which becomes flat at large field values:
\be
\label{eq:genpotential}
\tilde V(\chi) = \frac{\lambda }{4}\phi(\chi)^4\left(1+\xi\frac{\phi(\chi)^2}{M_P^2}\right)^{-2}\,,
\ee 
where $\phi$ parametrizes the one-dimensional inflaton direction before canonical normalization. $\lambda$ and $\xi$ in \eqref{eq:genpotential} represent effective quartic and non-minimal couplings, which depend on the orientation of the inflationary valley. The previous effective potential is non-renormalizable, and the theory can be seen to have an associated cutoff $\Lambda=M_P/\xi$ \cite{Burgess:2009ea,Barbon:2009ya}. On the other hand, CMB observations impose $\xi\sim 10^5 \sqrt{\lambda}$, and successful inflation requires values of the field $\chi$ up to $M_P/\sqrt{\xi}$. Note how in order to have field values below the cutoff during inflation, one needs $\xi\lesssim 1$. The direction of the inflationary valley  is determined by the signs of the parameter combinations $
\kappa_H  \equiv \lambda_{H\sigma} \xi_H - \lambda_H \xi_\sigma\,,
\kappa_\sigma  \equiv \lambda_{H\sigma} \xi_\sigma - \lambda_\sigma \xi_H.$ For inflation along the $H$ direction, one recovers the $H$ inflation scenario \cite{Bezrukov:2007ep} with $\lambda=\lambda_H, \xi=\xi_H$. $H$ mass measurements imply $\lambda_H=O(0.1)$, so that CMB constraints enforce $\xi_H\sim10^4$. This violates the $\xi\lesssim 1$ cutoff condition, which makes the predictions of $H$ inflation questionable. Thus we center on inflation along the $\sigma$ direction (HSI) or the mixed $H/\sigma$ direction (HHSI). Assuming $1\gtrsim\xi_\sigma\gg\xi_H$, for HSI one needs to use equation \eqref{eq:genpotential} with 
$\lambda=\lambda_\sigma,\xi=\xi_\sigma$, while for HHSI one has $\lambda=\lambda_\sigma-\lambda^2_{H\sigma}/\lambda_H,\xi=\xi_\sigma$. Then predictive inflation with $\xi\lesssim1$ (which can be seen to require $\lambda_\sigma\lesssim10^{-10}$) is compatible with CMB measurements, as seen in figure \ref{fig:CMB}, taken from \cite{Ballesteros:2016xej}, which shows the tensor-to-scalar-ratio $r$ and the spectral index $n_s$ at a reference scale $k_*=0.002{\,\rm Mpc}^{-1}$, together with the 68\% and 95\% confidence-level regions of Planck/BICEP \cite{Array:2015xqh}. The narrow SMASH predictions in the thick black line of figure \ref{fig:CMB} could be ruled out by future experiments such as CORE\cite{Finelli:2016cyd} and LiteBIRD\cite{Matsumura:2013aja}. One has $r\gtrsim0.004$, and it can also be seen that $\alpha=dn_s/d(log k_*)\gtrsim -8\times10^{-4}$.

In SMASH, the $H$ and $\sigma$ directions can be absolutely stable while providing successful inflation. This is essentially due to the threshold stabilization mechanism \cite{Lebedev:2012zw,EliasMiro:2012ay}, which, whenever $\sigma$ gets a large VEV, implies a non-trivial matching between the SM $H$ quartic $\overline{\lambda}_H$ and that of SMASH: $\lambda_H=\overline{\lambda}_H+\lambda^2_{H\sigma}/\lambda_\sigma$. This not only makes $\lambda_H$ larger than in the SM, but also enhances the  contributions of $\lambda_H$ to its beta function. This  results in enhanced stability in pure $H$ or mixed $H/\sigma$ directions. Stability can be seen to require $\delta\equiv \lambda^2_{H\sigma}/\lambda_\sigma$ in between $10^{-3}-10^{-1}$, depending on the top mass. Similarly, the $\sigma$ direction can remain stable despite the coupling of $\sigma$ to RH neutrinos, as long as they remain light enough ($\lesssim 10^7$ GeV for $\lambda_\sigma\lesssim10^{-10}$ and $v_\sigma\sim 10^{11}$ GeV, as preferred by predictive inflation and dark matter).
     
\subsection{Reheating}

After the slow-roll inflationary period, the effective inflaton field $\phi$ (aligned with $\sigma$ in HI, and mostly aligned with $\sigma$ with  a very small $H$ component in HHSI) oscillates in a potential dominated by a quartic interaction, behaving as a radiation fluid. The absence of a transition from matter-domination to radiation-domination at early times (and the absence of the corresponding uncertainty in the transition time), simplifies the matching  between perturbations generated during inflation and their associated length scales today, resulting in the narrow  SMASH predictions in figure \ref{fig:CMB}. During the oscillation phase, the background scalar field behaves as a condensate of particles with energies equal to the oscillation frequency,  $\omega\sim\sqrt{\lambda_{\sigma}}\phi_0$, with $\phi_0\sim M_P/a(t)$ corresponding to the amplitude of oscillations. The condensate can annihilate or decay into SM particles through multi-body processes, reheating the Universe; the production of bosonic particles can be enhanced by resonant effects. The $\phi\sim\sigma$ background gives a mass $m_h\sim\sqrt{\lambda_{H\sigma}}|\phi|$ to  $H$. Since predictive inflation and stability enforce $\lambda_{H\sigma}\gg \lambda_\sigma$, one typically has $m_h\gg\omega$, except very near the times at which the inflaton background crosses zero, making $m_h=0$. This makes $H$ production from the background very inefficient, while, on the other hand, $\sigma$ fluctuations can grow very fast thanks to unhindered resonant production. The large $\sigma$ fluctuations end up erasing the oscillations of the $\sigma$ background and restoring the PQ symmetry, as the phase of $\sigma$ acquires random values. In the case of HSI, the inflaton can only  directly produce $H$ bosons. The large $\sigma$ fluctuations cause the induced $H$ mass to stop oscillating, so that $H$ production becomes  blocked until late times, when the $\sigma$ fluctuations become redshifted. As a consequence of this, the relative amount of energy stored in  the fluctuations of the phase of $\sigma$ (axions) ends up being too large, and HSI scenarios are ruled out by predicting a change in the effective number of relativistic neutrinos $\Delta N_{\rm eff}\sim1$, incompatible with the Planck result $N_{\rm eff}=3.04\pm0.18$ \cite{Ade:2015xua}. In HHSI scenarios, on the other hand, the inflaton has an $H$ component which can produce gauge bosons. As argued before, $H$ fluctuations are suppressed, so that the oscillations of the $H$ component of the condensate are not erased. Then the gauge bosons maintain an oscillating mass in the  background, and although this mass remains above $\omega$ away from the crossings, particle production is allowed during them. This ends up being enough to successfully reheat the Universe. The gauge fields decay efficiently into light fermions when acquiring a mass in between crossings,  and the light particles eventually thermalize. This allows for an additional thermal source of gauge boson production at the crossings. The new gauge bosons steal energy from the condensate as they acquire a mass away from the crossings and decay back to the SM plasma. This thermal feedback mechanism enhances the rate of energy loss from the scalar condensate and allows for efficient reheating. We estimate reheating temperatures of the order of $T_R\sim 10^{10}$ GeV in HHSI.

\subsection{PQ breaking, baryogenesis and neutrino masses}

The high reheating temperature  $T_R\sim 10^{10}$ GeV in HHSI is enough to give rise to a thermal restoration of the PQ symmetry. Since the RH neutrinos can only acquire a mass through the VEV of $\sigma$, they 
remain massless in the early Universe, and can achieve thermal equilibrium abundances. As the Universe cools down, the PQ symmetry is broken at $T_c\sim \lambda^{1/4}_\sigma v_\sigma $, which for the preferred values $\lambda_\sigma\sim 10^{-10}, v_\sigma\sim 10^{11}$ GeV is of the order of $T_c\sim10^8$ GeV. As argued in section \ref{subsec:inflst}, for the previous choice of parameters stability demands $M_i\lesssim10^7$ GeV; thus, after the PQ phase transition the temperature is still much larger than the masses of the $N_i$, which can then retain an equilibrium abundance. As the temperature drops below the mass of the $N_i$, the inverse decays fall out of equilibrium and the $N_i$ can decay into $H$ bosons and leptons. The baryon asymmetry can be generated through leptogenesis \cite{Fukugita:1986hr}: the $N_i$ decays give rise to an asymmetry in lepton number, which is reprocessed into a baryon asymmetry by SU(2) sphaleron interactions (transitions between topological vacua). In SMASH, the $N_i$ can also annihilate into $\sigma$ without contributing to an asymmetry, but the rate is very suppressed with respect to decays and can be ignored. Scenarios with hierarchical $N_i$ masses require the lightest mass to satisfy $M_1\lesssim 5\times10^8$ GeV \cite{Davidson:2002qv,Buchmuller:2002rq}, but the bound can be circumvented with some degree of degeneracy between the $N_i$ \cite{Pilaftsis:2003gt}. For masses of the ${N_i}$ near $10^7$ GeV, as required by stability and dark matter, one needs a degeneracy around $4\%$, much milder than in the usual resonant leptogenesis scenarios.
Finally, as the Universe keeps cooling down and the electroweak symmetry becomes broken, the light neutrinos acquire small masses  $m_\nu\sim v^2/v_\sigma F Y^{-1}F^\top$. 

\subsection{Dark matter and the strong CP problem}

The phase of the field $\sigma$, which contains the axion field $A$ (${\rm Arg}\,\sigma= A/v_\sigma$), acquires couplings to the SU(3) and U(1) pseudoscalar densities (due to the PQ anomalies under both gauge groups) as well as derivative couplings to SM fermions.
The large fluctuations of $\sigma$ during the preheating phase give rise to an excited axion condensate whose energy redshifts as the Universe cools down. The reheating temperature of the plasma, $T_R\sim 10^{10}$ GeV, is enough to produce an additional thermal population of higher-energy axion excitations thanks to the anomalous/derivative couplings with the SM fields. This relativistic population freezes at a temperature around $10^9$ GeV, at which the SM interactions of the axion decouple. 
When the PQ symmetry becomes broken, the $|\sigma|$ fluctuations localize around the corresponding VEV, while the axion condensate retains random values in different patches of the Universe, 
separated by domain walls which intersect at axion strings. At temperatures at which QCD interactions become strong, the axion condensate develops a mass as a consequence of its anomalous QCD coupling. Due to this,  the displaced axion condensate starts oscillating around zero in the different patches of the Universe, with random initial values. This dynamics, known as the ``misalignment mechanism'', makes the condensate behave as a non-relativistic fluid, acting as dark matter. On the other hand, axion domain walls and strings are unstable, since  the choice of matter in SMASH, enforcing a particular form of the PQ anomaly, implies that each string is only attached to a single domain wall. This allows strings to reconnect in loops which can subsequently decay by emitting axion excitations and gravitational waves. The axion production by strings is dominated by the late time contributions, when the emitted axions are also non-relativistic and behave as dark matter. The dark matter contribution from the misalignment mechanism can be understood precisely thanks to recent advances in lattice calculations of the finite-temperature axion mass due to QCD effects \cite{Borsanyi:2016ksw}; the contribution from decaying strings is more uncertain. Within the uncertainty, the requirement of fitting the observed dark matter relic abundance with axions imposes $3\times 10^{10}{\,\rm GeV}\lesssim v_\sigma\lesssim 1.2\times 10^{11}{\,\rm GeV}$. This in turn enforces $50\mu{\rm eV}\lesssim m_A\lesssim 200 \mu{\rm eV}$, which could be probed by future experiments such as CULTASK \cite{Woohyun:2016hkn}, MADMAX \cite{TheMADMAXWorkingGroup:2016hpc} and ORPHEUS \cite{Rybka:2014cya}. As pertains to the decoupled population of relativistic axions originated from the hot SM plasma, it contributes to the effective number of relativistic neutrinos by an amount fixed by
the thermal equilibrium at early times; this gives $\Delta N_{\rm eff}\sim 0.03$, which could be probed by future CMB polarization measurements \cite{Abazajian:2013oma,Errard:2015cxa}. Finally, since the anomalies enforce axions to enter the effective action in the physical combination $ \theta_{\rm phys}(x)\equiv\theta+A(x)/v_\sigma$, the QCD-generated axion mass implies that $\langle\theta_{\rm phys}(x)\rangle=0$. As $\theta_{\rm phys}$ is the only combination that can enter CP-violating  observables such as the neutron's dipole moment, this solves the strong CP problem.

\section{Conclusions}

We have provided an overview of SMASH, an extension of the Standard Model with right-handed neutrinos, a  complex singlet scalar and a vector-like quark, which features a single new mass scale $v_\sigma\sim10^{11}$ GeV and provides a falsifiable framework that  addresses the following problems in particle physics and cosmology: inflation, $H$ stability, baryogenesis, neutrino masses, dark matter, and the strong CP problem. The theory predicts a tensor-to-scalar-ratio $r\gtrsim0.004$, a running of the spectral index $\alpha\gtrsim-8\times 10^{-4}$, and a deviation in the effective number of relativistic neutrino species $\Delta N_{\nu\rm eff}\sim0.03$. These values can be probed in future CMB experiments, such as LiteBIRD and CORE. The model predicts an axion in the mass window $50\mu{\rm eV}\lesssim m_A\lesssim 200 \mu{\rm eV}$, in the reach of future axion experiments such as CULTASK, MADMAX and ORPHEUS.

\section*{Acknowledgements}

Many thanks to Guillermo Ballesteros, Andreas Ringwald and Javier Redondo for a smashing collaborative experience. And very, very special thanks to the Moriond secretaries and IT staff for going beyond their duty to help me in times of self-inflicted, computerless need.

\end{document}